\documentstyle[aps,prb]{revtex}

\input{epsf}

\begin{document}

\title{Magnetic properties of a two-electron quantum dot}
\author{C~E~Creffield$^{1,3}$, J~H~Jefferson$^2$, Sarben Sarkar$^3$ 
and D~L~J~Tipton$^{2,3}$}

\address{
$^1$Instituto de Ciencia de Materiales (CSIC), Cantoblanco, E-28049, Madrid,
Spain \\
$^2$Defence Evaluation and Research Agency, Electronics Sector, St. Andrews\\
Road, Malvern, Worcs. WR14~3PS, UK \\
$^3$Dept. of Physics, King's College London, Strand, London, WC2R~2LS, UK}

\date{\today}

\maketitle

\begin{abstract}
The low-energy eigenstates of two interacting electrons in a square quantum
dot in a magnetic field are determined by numerical diagonalization. In the
strong correlation regime, the low-energy eigenstates show Aharonov-Bohm
type oscillations, which decrease in amplitude as the field increases. These
oscillations, including the decrease in amplitude, may be reproduced to good
accuracy by an extended Hubbard model in a basis of localized one-electron
Hartree states. The hopping matrix element, $t$, comprises the usual kinetic
energy term plus a term derived from the Coulomb interaction. The latter is
essential to get good agreement with exact results. The phase of $t$ gives
rise to the usual Peierls factor, related to the flux through a square
defined by the peaks of the Hartree wavefunctions. The magnitude of $t$
decreases slowly with magnetic field as the Hartree functions become more
localized, giving rise to the decreasing amplitude of the Aharonov-Bohm
oscillations.
\end{abstract}

\bigskip

Advances in the fabrication of semiconductor nanostructures have made it
possible to construct devices, termed ``quantum dots'', in which a small
number of electrons can be confined to regions of the order of the Fermi
wavelength. Due to this confinement the electronic spectrum of a quantum dot
is composed of discrete levels, which have been studied in detail by
conductance and spectroscopic measurements. Interactions between electrons
have been shown to be of major importance in determining the electronic
properties of these systems, and the effects of the strong correlations
between particles has attracted intense experimental and theoretical
attention. A particular motivation for studying the properties of few
electron quantum dots is their relevance to the rapidly developing field of
quantum computing [\cite{loss}], as the entangled states of the electrons
confined in a quantum dot can give a physical realization of a quantum bit
or ``qubit''. A convenient probe to study quantum dots is the application of
a magnetic field which has revealed numerous dramatic and novel quantum
effects, such as parity oscillations of the ground-state [\cite{wagner}] and
the phenomenon of magic numbers [\cite{maksym}]. A variety of 
techniques have been developed to treat these systems, including numerical
QMC [\cite{bolton},\cite{egger}], Hartree-Fock diagonalization 
[\cite {pfannkuche}], and the direct diagonalization of the 
many-body Hamiltonian [\cite{maksym}]. 
The majority of these treatments, however, have concentrated
on circularly symmetric dots with a parabolic confining potential. This is
indeed a reasonable approximation to the potential found experimentally in
large dots produced by ``soft'' confinement, and it is appealing from the
theoretical point of view because the rotational symmetry of the potential
renders the single-particle problem completely integrable [\cite{fock}, 
\cite{darwin}], giving a natural basis to describe the many-particle
situation. In real devices, however, one must expect deviations from perfect
symmetry, which can have profound effects on their spectrum [\cite{gudmund}],
and furthermore for dots created by ``hard'' confinement, such as a
heterojunction between semiconductors, the parabolic approximation is
unrealistic as the potential is essentially flat within the dot and sharply
rising at the dot boundary. A square-well bounded by infinite potential
barriers provides a simple, though idealized, model of this form of
confinement. In contrast to rotationally symmetric potentials, angular
momentum is not a good quantum number for the square-well system except in
the limit of very high fields, and accordingly even the single-particle
problem is non-integrable. Another consequence of the square symmetry is
that this system can exhibit many-body effects which are not readily
observed in parabolically confined dots [\cite{ugajin}], as the boundary
conditions mean that the spatial co-ordinates cannot be split into center of
mass and relative co-ordinates, and hence Kohn's theorem is not applicable.\\

In this paper we consider the case of two electrons confined to a
square-well quantum dot in the presence of a magnetic field. In a
square-well potential the kinetic energy scales like $1/L^{2}$, and the
interaction energy like $1/L$ where $L$ is the side-length of the well. The
competition between these terms determines the nature of the electron
system. For small $L$ the Coulomb energy is insignificant in comparison to
the single-particle kinetic energies and the electrons behave like
uncorrelated independent particles. As $L$ increases the Coulomb term
becomes increasingly dominant and the many-particle states become correlated
and cannot be described in terms of an independent electron picture. In this
limit the electrons will form a quasi-crystalline state
to minimize their electrostatic energy, termed a ``Wigner molecule'
in analogy with the
formation of a Wigner crystal in an infinite system [\cite{wigner}]. This
observation is the key to a method developed to treat the strongly
correlated situation by mapping the low energy spectrum of the system to an
effective lattice model of the Hubbard type [\cite{jhjprb}], where the
lattice points are identified with the peaks in the charge density of the
Wigner molecule state. In the absence of a magnetic field this technique has
been successfully used to treat one-dimensional quantum dots containing up
to six electrons [\cite{jhjprb}], and also for two-dimensional polygonal
dots containing two electrons [\cite{prb}]. If this mapping to a lattice
model indeed captures the basic physics of the system, we should expect it
to also reproduce the dominant effects of an applied magnetic field by the
the inclusion of a Peierls factor [\cite{peierls}] in the inter-site hopping
terms. We test this expectation here by comparing the predictions of the
lattice model modified in this way with results obtained by the explicit
diagonalization of two-electron quantum dots pierced by a variety of
different magnetic flux distributions. We find that in all cases the Peierls
substitution gives the correct qualitative behavior of the energy levels,
and becomes precise in the limit of an Aharonov-Bohm flux-line. Quantitative
accuracy for physically realizable magnetic fields may be obtained by minor
renormalizations of the parameters of the effective lattice model, arising
from the increased localization of the electron wavefunctions produced by a
physical magnetic field. We further justify this picture by explicitly
constructing localized Hartree basis functions. The Hamiltonian is then
diagonalized using the lowest four Hartree basis functions, corresponding to
an electron localized near one of the four corners of the square. This gives
good agreement with the exact results. Furthermore, retaining only
nearest-neighbor hopping and diagonal Coulomb repulsion terms also gives
good quantitative agreement with the exact result, provided that the hopping
term includes an important contribution from the Coulomb interaction. This
gives explicit justification of the simple extended Hubbard model, 
and correctly reproduces the detailed dependence of the energy levels on
the applied field. Finally we will discuss the relevance of these
results to the phenomena predicted to occur in quantum dot arrays 
[\cite{stafford}], which can also be described by a mapping to a lattice 
model of Hubbard type.\\

We consider a square quantum dot, described by the Hamiltonian: 
\begin{equation}  \label{ham}
H = \frac{1}{2 m^{\ast}} \sum_{i=1}^{2} 
\left[ (-i \hbar \nabla_i + e{\bf A}_i)^2 + V({\bf r}_i) \right] \ + \ 
\frac{e^2}{4 \pi \epsilon_0 \epsilon_r} 
\frac{1}{|{\bf r}_1 - {\bf r}_2 |}
\end{equation}
where $V({\bf r})$ is the confinement potential, and ${\bf A}$ is the vector
potential of the applied magnetic field ${\bf B}$. We assume that the
electrons are confined by infinite walls, and that they can be described by
the effective mass approximation. The two simplest forms of applied magnetic
field are a uniform field, $B_z$, perpendicular to the plane of the dot,
and, at the other extreme, an Aharonov-Bohm flux-line in which the magnetic
field is zero throughout the dot except at a single point, but phase
interference effects arise from the particles' interactions with the vector
potential. We may interpolate between these cases by using a flux-tube of
radius $a$, inside which the magnetic field is uniform, and
outside of which the field is zero. This is equivalent to the field produced
by an infinitely long solenoid of radius $a$. In the symmetric gauge it can
easily be shown that the vector potential corresponding to this field is
given by: 
\begin{equation}
{\bf A}=(A_r, A_{\theta}) = \left\{ 
\begin{array}{r@{\quad:\quad}l}
\left( 0, \frac{B r}{2} \right) & \mbox{for} \ r \leq a \\ 
\left( 0, \frac{B a^2}{2 r} \right) & r > a
\end{array}
\right.  \label{flux}
\end{equation}
However, to simplify the numerical investigation it was decided to use an
analytic form for the vector potential: 
\begin{equation}
A_{\theta} = \frac{B a^2}{2} \frac{r}{r^2 + a^2}
\end{equation}
which has the same behavior as (\ref{flux}) at small and large values of 
$r$, but does not have a cusp at $r = a$ and is therefore more amenable to a
numerical treatment. Aligning the flux-tube with the center of the dot
allows us to easily compare the results obtained as the magnetic field is
changed from the uniform case ($a \rightarrow \infty$) toward the limit of
an Aharonov-Bohm flux-line ($a=0$) by altering the value of $a$.\\

To find the eigenvalues of the Hamiltonian (\ref{ham}) we chose to use a
basis of states $\{\psi_n(x) \psi_m(y)\}$ for each electron, where: 
\begin{equation}
\psi_n(x) = \sqrt{\frac{2}{L}} \ \sin(\frac{n \pi x}{L}) \ , \quad 0 \leq x
\leq L
\end{equation}
and $L$ is the side-length of the square dot, which are the eigenstates of
the non-interacting system in the absence of a magnetic field. For the
specific case of a uniform magnetic field the matrix elements of all the
single-electron terms in the Hamiltonian (\ref{ham}) can be calculated
analytically in this basis. For the other field configurations, however, it
was necessary to calculate these matrix elements numerically, and a NAG
routine was used to evaluate the two-dimensional integrals. The matrix
elements of the Coulomb interaction were also obtained numerically.
Up to eight basis functions per direction were used for each particle 
to obtain the matrix form of the Hamiltonian, which was then
block-diagonalized into singlet and triplet subspaces and treated by a
standard eigenvalue routine. In all cases the dot material was taken to be
GaAs, with an effective mass $m^{\ast} = 0.067 m_e$, and a relative
permittivity $\epsilon_r = 10.9$, giving an effective Bohr radius of $a_B =
8.8$nm.\\

\begin{figure}[tbh]
\centerline{\epsfxsize=100mm \epsfbox{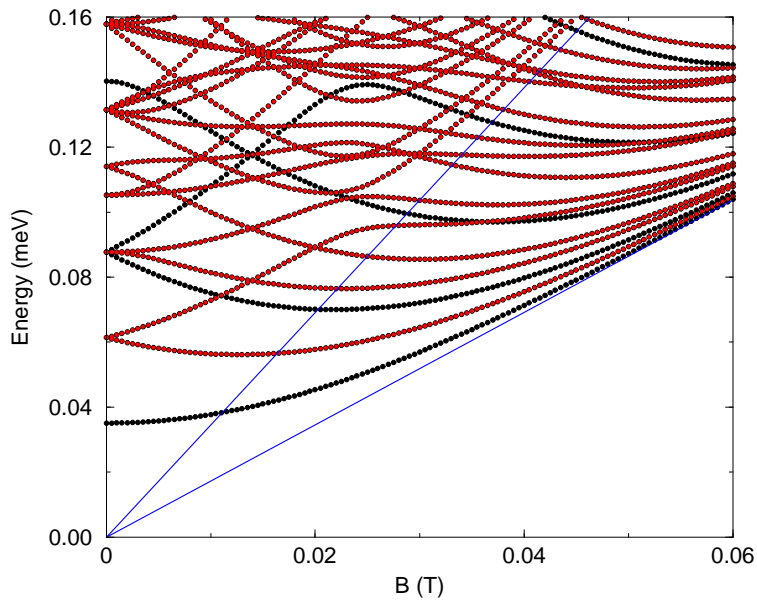}}
\caption{\label{non-int}
Energy levels of two non-interacting electrons in a $800$nm square-well quantum
dot. Black and red lines indicate singlet and triplet states respectively.
The solid blue lines indicate the lowest two Landau levels.}
\end{figure}

We first consider the case of non-interacting electrons in the presence of a
uniform magnetic field in order to understand the single-particle effects of
the magnetic field, before introducing the additional complication of strong
correlations. We show in Fig.[\ref{non-int}] the dependence of the energies
of the low-lying eigenstates on the strength of the applied field $B_z$ for
the two non-interacting electrons. At low field strengths the spectrum is
extremely complex, with some levels showing a linear dependence on the
field, and others a quadratic dependence. Many level crossings occur, and
also the phenomenon of of ``avoided crossings'' or level repulsion can be
clearly seen. This is a typical signature of the presence of quantum chaos,
and arises here because of the non-integrability of the single-particle
problem in a square boundary. Such avoided crossings are not seen, for
example, when the confining boundary is rotationally symmetric. Avoided
crossings have also been seen in experiments, such as in Ref.[\cite{expt}],
in which electron micrographs of the quantum dots clearly show deviation
from circular symmetry. At high magnetic fields the spectrum simplifies
considerably, and it can be seen that the energy levels start to condense
into highly degenerate Landau levels with a linear dependence on the field, 
$E_n = (n + 1) \hbar \omega_c$, where $\omega_c = e B/m^{\ast}$ is the
cyclotron frequency. It should be noted that the ground-state experiences no
level crossings and evolves smoothly into the lowest Landau level, remaining
a spin singlet for all values of magnetic field.\\

\begin{figure}
\hspace*{-1 cm}
\epsfxsize=20 cm
\epsfysize=9 cm
\epsffile{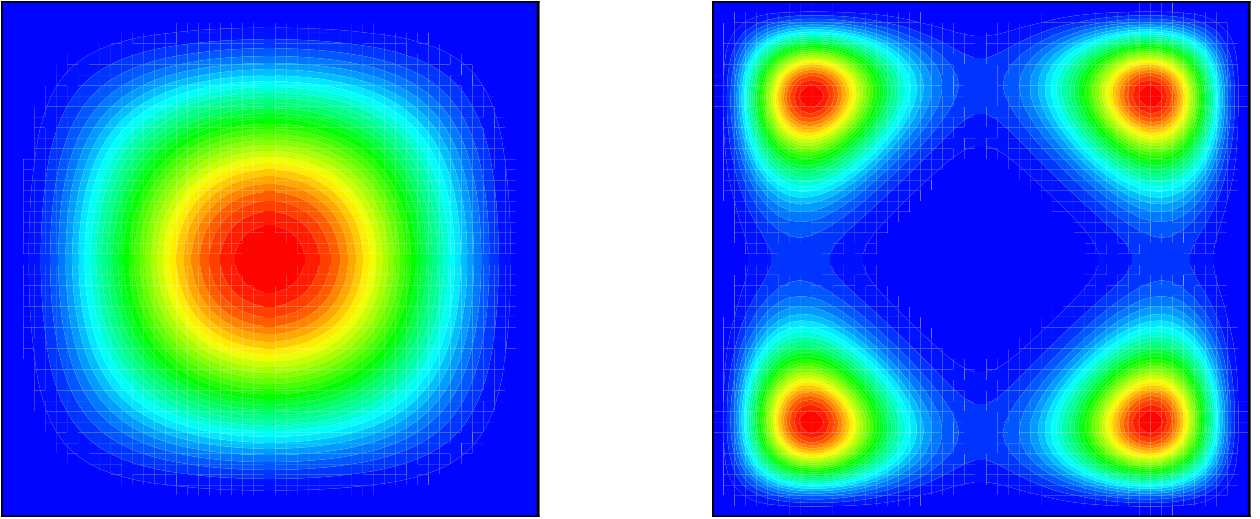}
\caption{\label{squares}
Ground state charge density distributions for a) $L=10$nm, 
b) L=$1600$nm showing the transition from a weakly interacting case to 
a sharply peaked Wigner crystal. These peaks define the lattice points 
used in the effective charge-spin model (see text).}
\end{figure}

We now consider the effect of turning on the Coulomb interaction. As was
stated earlier, the physical size, $L$, of the dot determines the relative
importance of the kinetic and Coulomb energies of the system, and when the
mean electron separation exceeds a critical value $r_{c}$ the electron
charge density becomes localized in space, forming a Wigner molecule. It was
found in Ref.[\cite{prb}] that for a two-electron polygonal dot 
$r_{c}\approx 10a_{B}$. We show in Fig.[\ref{squares}] the ground-state
charge densities for two extreme cases $L\ll r_{c}$ and $L\gg r_{c}$ to show
how the structure of the ground-state alters as $r_{c}$ is exceeded. The
existence of the strongly-correlated Wigner molecule state is the criterion
for the validity of mapping the low-lying states of the quantum dot to an
effective lattice model, where the lattice sites are given by the location
of the peaks of the Wigner molecule state. In [\cite{jhjprb}] it was
conjectured that the appropriate effective lattice model to describe the
low-energy manifold of a system of strongly interacting electrons in a
quantum dot is an extended single-band Hubbard model. If, for this case of
two electrons, we neglect direct exchange and set the Hubbard-$U$ energy to
infinity (equivalent to forbidding double occupation of `lattice' sites and
neglecting superexchange) then the extended Hubbard model takes on the
particularly simple form: 
\begin{equation}
H^{tV}=E_{0}+{\cal P}\left[ \sum_{\langle i,j\rangle \sigma }\left(
t c_{i\sigma }^{\dagger }c_{j\sigma }+\mbox{h.c.}\right) +Vn_{i}n_{j}\right] 
{\cal P}.  \label{tv}
\end{equation}
Here ${\cal P}$ is a projection operator which eliminates doubly-occupied
lattice sites, $t$ is a nearest neighbors hopping term and $V$ is
the difference in Coulomb energy between states when the two electrons  
occupy neighboring sites and when they are diagonally opposite each other.

The energy $E_{0}=2\varepsilon_{0}+V_{0}$ 
is the ground-state energy in the `atomic limit', 
where $\varepsilon_{0}$ is the on-site energy and $V_{0}$ is the
Coulomb energy between the two electrons on diagonally opposite sites. The
energy parameters $E_{0}$, $V$ and $t$ depend on the magnitude of the
magnetic field, and $t$ in general is complex. One can see on inspection that
the spectrum of the Hamiltonian (\ref{tv}) must consist of a low-lying
manifold of two singlets and two triplets, corresponding to the electrons
being on diagonally opposite sites, and a higher lying manifold consisting
of four singlets and four triplets, corresponding to the two electrons being
on neighboring sites. Furthermore, in the strong correlation regime for which 
$V \gg t$, the lower manifold should have energy $\sim E_{0}$ with the
higher-lying manifold at energy $\sim E_{0}+V$. This is indeed the case as
can be seen from the exact analytic solutions which take the form: 
\begin{equation}
E_{{\rm s}}-E_{0}=\frac{1}{2}\left[ {V\pm \sqrt{V^{2}+32|t|^{2}\sin
^{2}\phi }}\right], \ \frac{1}{2}\left[ {V\pm \sqrt{V^{2}+32|t|^{2}\cos
^{2}\phi }}\right], \ V, \ V  \label{singlets}
\end{equation}
for the singlets and 
\begin{equation}
E_{{\rm t}}-E_{0}=\frac{1}{2}\left[ {V\pm \sqrt{V^{2}+16|t|^{2}(1+\sin
2\phi })}\right], \ \frac{1}{2}\left[ {V\pm \sqrt{V^{2}+16|t|^{2}(1-\sin 2\phi 
})}\right], \ V, \ V  \label{triplets}
\end{equation}
for the triplets, where we have set $t=|t|e^{i\phi }$.
Since $|t| \ll V$, we may expand the square root in equations (\ref{singlets})
and (\ref{triplets}). To second-order this gives for the lowest manifold of
states: 
\begin{eqnarray}
E_{{\rm s}} &=&\tilde{E}_{0}\pm 2\Delta \cos 2\phi  \nonumber \\
E_{t} &=&\tilde{E}_{0}\pm 2\Delta \sin 2\phi  \label{delta}
\end{eqnarray}
where $\Delta =2|t|^{2}/V$ and $\tilde{E}_{0}=E_{0}-2\Delta $. We can, in
fact, derive this result in a different way, which emphasizes the nature of
the low-lying eigenstates. The excited states corresponding to the two
electrons on neighboring sites are eliminated by degenerate perturbation
theory. To second-order, the $tV$-Hamiltonian is then transformed 
into the effective Hamiltonian \cite{jhjprb,prb}: 
\begin{equation}
H_{{\rm eff}}=\tilde{E}_{0}+
(\Delta e^{i2\phi }R_{\frac{\pi }{2}}+{\rm h.c.})  \label{heff}
\end{equation}
where $R_{\frac{\pi }{2}}$ rotates the two electrons at opposite corners of
the square on a diagonal by $\pi /2$. Diagonalization of $H_{{\rm eff}}$
yields directly the singlet and triplet energies (\ref{delta}). In this
effective model the pair of electrons thus tunnel between the base states
with an amplitude modulated by a Peierls factor, $e^{i2\phi }$, with twice
the phase angle of the underlying extended Hubbard model, since it involves
two electron hops. Although $H_{{\rm eff}}$ reproduces the low-energy
multiplet of (\ref{tv}) to good accuracy, it is instructive to calculate the
next (fourth-order) correction. This renormalizes $\tilde{E}_{0}$ and 
$\Delta$, and also introduces a Heisenberg spin exchange term: 
\begin{equation}
J\left[ {\bf s}_{1}\cdot {\bf s}_{3}+{\bf s}_{2}\cdot {\bf s}_{4}\right] 
\end{equation}
where $J=-\frac{32|t|^{4}}{V^{3}}\cos 4\phi $ (ferromagnetic). Although this
is a small correction, it is not negligible and accounts, for example, for
the small asymmetry of the singlet-triplet splittings at $B=0$. The
fourth-order effective Hamiltonian has, in fact, the most general form, since
higher orders cannot introduce qualitatively different terms. Thus we see
that the low-energy manifold can be generally described by a charge-spin
model in which the electrons rotate rigidly and undergo Heisenberg exchange.\\

\begin{figure}
\centerline{\epsfxsize=100mm \epsfbox{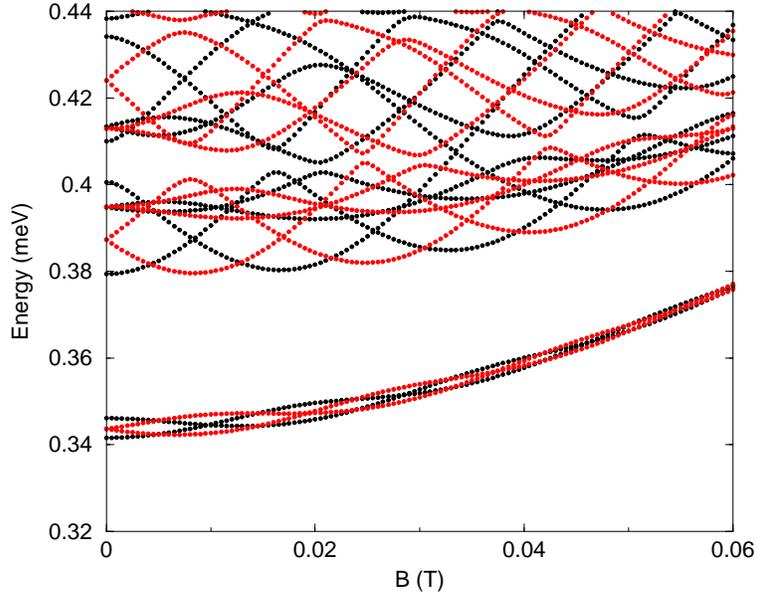}}
\caption{\label{uniform}
Energy levels of a two electron square-well quantum dot in a
uniform magnetic field. Black and red lines indicate singlet and triplet
states respectively.}
\end{figure}

\begin{figure}
\centerline{\epsfxsize=100mm \epsfbox{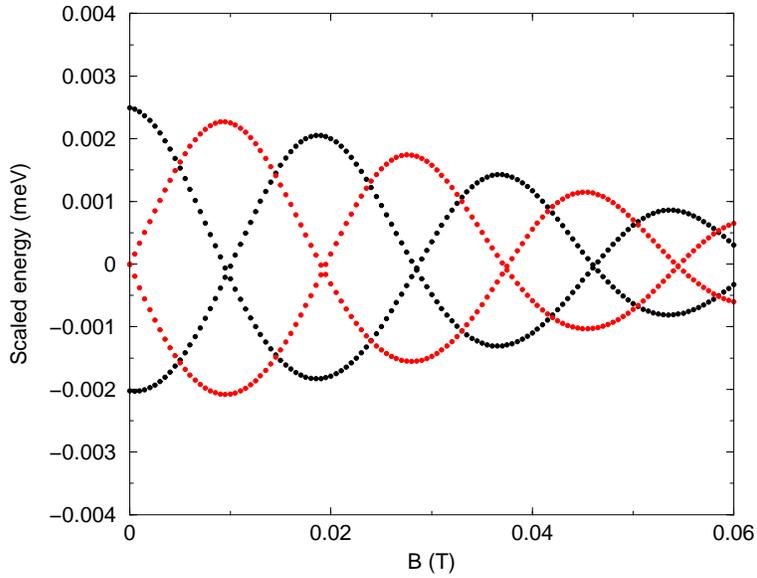}}
\caption{\label{oscs}
Dependence of the lowest lying energy levels on magnetic field,
with the average increase subtracted.}
\end{figure}

In Fig.[\ref{uniform}] we show the evolution of the lowest energy levels as
a function of the field strength of a uniform magnetic field. The dot-size
was taken to be $L=800$nm, which is well within the Wigner molecule regime.
We see that the structure of the lowest multiplet agrees with the above
prediction, and that this multiplet is relatively isolated from the next
highest set of states due to the size of the $V$ term. The Zeeman term has
been neglected in the initial Hamiltonian (\ref{ham}). This would, of
course, produce small splittings between the constituent states of each
triplet. In order to make a quantitative comparison of
the lowest multiplet with the form (\ref{heff}), we first subtract the
overall increase in energy of the ground state multiplet,
which depends approximately quadratically on the
applied field. The fine details of the oscillations in energy are then seen
clearly, as shown in Fig.[\ref{oscs}].\\

It is easily verified that (\ref{heff}) has the correct form to describe the
pattern of oscillations seen in Fig.[\ref{oscs}] with the quantity $\phi$
being the phase acquired by an electron hopping between two adjacent sites
of the lattice on which the effective $tV$ model is defined.
The total phase angle acquired when an electron hops once around the square
is $4\phi =2\pi \Phi /\Phi _{0}$, where $\Phi _{0}=h/e$ is the fundamental
flux quantum and $\Phi =Br_{L}^{2}$ is the total enclosed flux for an
effective lattice parameter $r_{L}$. By measuring the distance between the
peaks in the ground-state charge distribution we obtain a value of 
$r_{L}=495 $nm for an $800$nm dot, and thus a magnetic field of $0.0095$T
corresponds to a magnetic flux of $1.11\Phi _{0}/2$, which agrees well with
the value of $\Phi _{0}/2$ which Eqs.(\ref{delta}) predict for the form of
the spectrum to become inverted. This close agreement, obtained with {\em no}
adjustable parameters, fully endorses interpreting the results at non-zero
magnetic field within the framework of the lattice model in which the
hopping terms are modified by phase factors. 
An alternative interpretation of this behavior is that the two electrons
are hopping around the four sites of the square, giving rise to a persistent
current $I=-\partial E/\partial \Phi $. This is periodic in the enclosed
flux but with half the period of two non-interacting electrons on a ring,
due to the correlated positions of the rotating electrons, described by 
Eq. (\ref{delta}) It is interesting to note that the inverted spectrum which
occurs after a quarter cycle is the result that would be obtained for a 
{\em bosonic} $tV$ model in zero field [\cite{jhjprb}]. This is due to the 
fact that at this value of flux the Aharonov-Bohm phase acquired by the 
electrons when they exchange positions compensates for the sign arising 
from their fermionic statistics, and hence the magnetic field can be viewed 
as converting the electrons into ``composite bosons''.\\

A consequence of the oscillations in energy of the singlet and triplet
states in the lowest multiplet is that the parity of the ground-state
periodically changes from symmetric to antisymmetric. Similar parity
oscillations were observed by Wagner {\it et al} [\cite{wagner}] in the
spectrum of a two-electron quantum dot with a parabolic potential, and also
by Ugajin [\cite{ugajin}] in a numerical study of square dots of small
size. As the ground-state does not flip parity in the non-interacting case,
this behavior is clearly a consequence of the Coulomb interaction.\\

Although the Peierls substitution accounts for the oscillatory behavior of
the lowest energy levels, the precise details of their
behavior such as the decay in amplitude of the oscillations, the
marginal increase in their period (apparent from a close examination of 
Fig.[\ref{oscs}]), and the overall increase of the energy levels
with the field require additional explanation. This may be achieved 
by deriving suitable localized basis functions, and explicitly calculating 
the energy parameters of the resulting tight-binding (extended Hubbard) 
model. We have shown that this gives high accuracy for a basis set 
constructed from Hartree functions in the following way. We first fix the 
position of one electron in one corner of the square. The one-electron 
Schr\"odinger equation is then solved for the other electron, giving a set 
of one-electron states. For the Wigner regime considered here, the low-lying 
states, and in particular the ground-state, are localized near the corner of 
the square diagonally opposite the fixed electron. We then solve the 
Schr\"odinger equation for the first electron under the influence of the 
probability charge-density of the other, in the sense of Hartree. This 
procedure is iterated to convergence, yielding two sets of one-electron 
wavefunctions which may be mapped into one another by a rotation of $\pi$. 
Clearly two further sets may be deduced for the remaining corners of the 
square by a rotation of $\pi/2$. Each of these four sets is composed of 
mutually orthogonal single-electron wavefunctions, but wavefunctions from 
different sets are not necessarily orthogonal to each other. A complete 
orthonormal set maybe constructed progressively by first mutually 
orthogonalizing the ground-state from each set using L\"owdin's 
[\cite{lowdin}] method. These states alone enable the extended Hubbard model 
to be constructed and, as we will show, reproduce the exact results to good 
accuracy. The remaining Hartree excited states may then be orthogonalized to 
these lowest states by the Schmidt procedure. Their effect may be accounted 
for later by perturbation theory, where, for the regime of 
interest, they give small corrections.\\ 

\begin{figure}
\centerline{\epsfxsize=100mm \epsfbox{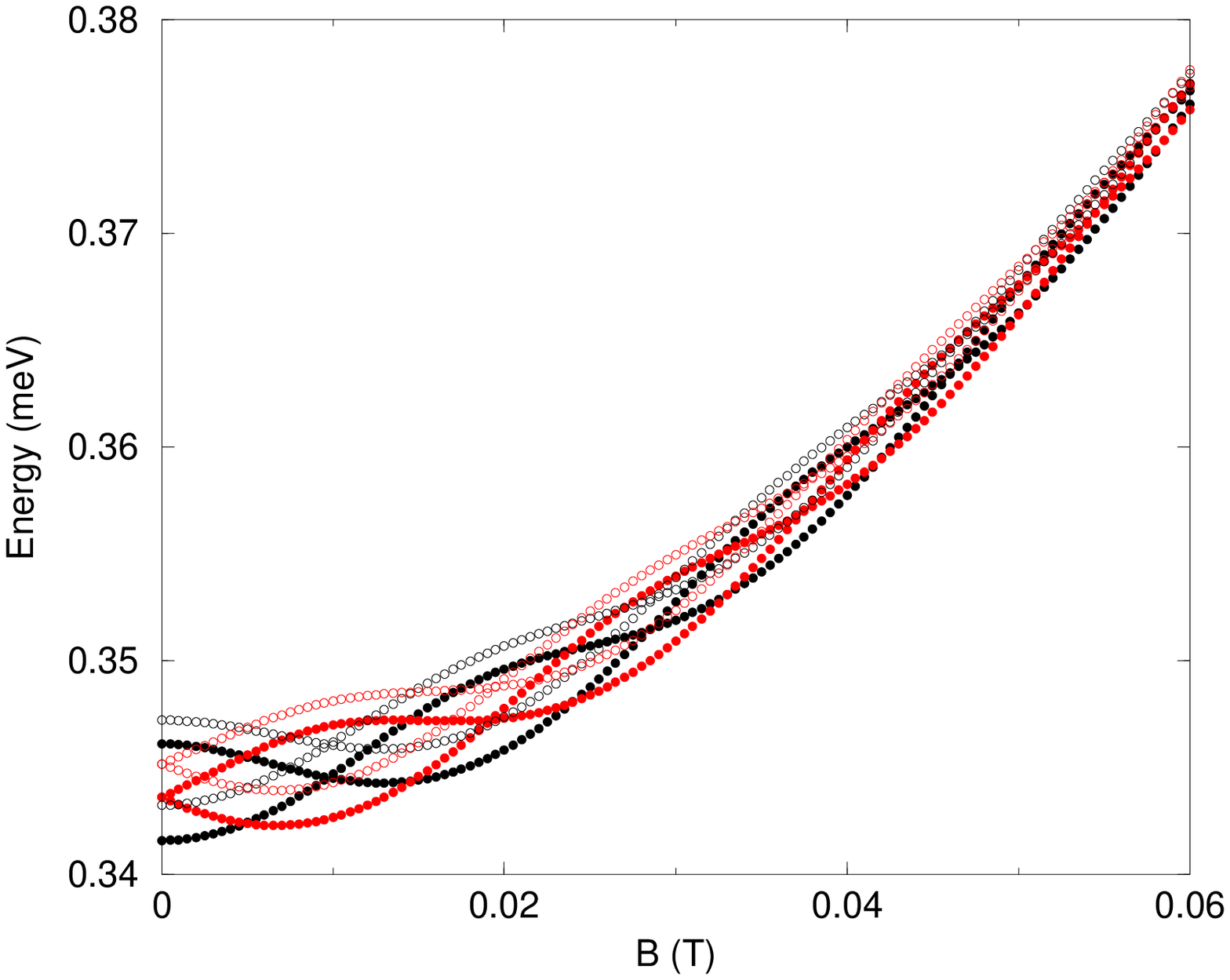}}
\centerline{\epsfxsize=100mm \epsfbox{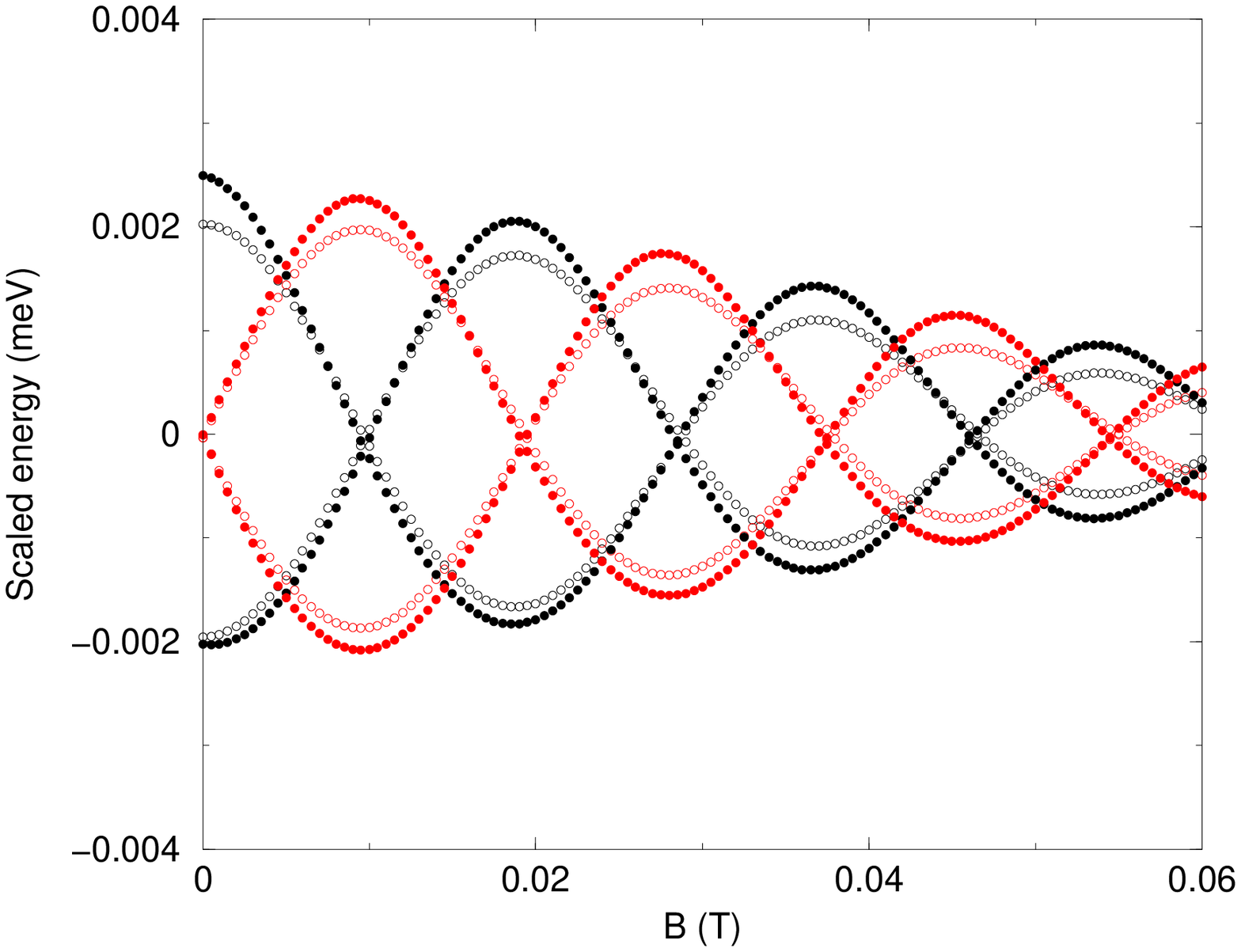}}
\caption{\label{hartree} Comparison of diagonalization within the space
of one Hartree wavefunction per site (open circles) and exact calculation 
(filled circles). The overall increase in energy with
field has been removed in the lower figure.}
\end{figure}

Restricting ourselves to the lowest four orthogonalized Hartree 
states we may construct all two-electron symmetric (singlet) and 
antisymmetric (triplet) orbital states and within this two-electron basis 
diagonalize the Hamiltonian (\ref{ham}). The results of this calculation for 
the lowest multiplet are shown in Fig.[\ref{hartree}] in which we also plot 
the exact results for comparison. We see that the agreement is good and all 
the essential physics is reproduced. In particular we see the overall 
increase in energy with magnetic field, and the corresponding decrease in 
amplitude of the oscillations. These features may be simply related to the 
change in shape of the one-electron Hartree functions, which become more 
localized with increasing magnetic field, thereby increasing one-electron 
energy whilst decreasing overlaps. We emphasize the importance of using 
Hartree rather than Hartree-Fock base states. Whilst the latter will always 
give the lowest estimate of the ground-state energy for a single Slater 
determinant, it does not necessarily (or even usually) give a very accurate 
estimate of a low-lying multiplet when there is near degeneracy. This is 
indeed the case for two electrons in a square dot. If we take the lowest 
four Hartree-Fock states (which are of course orthogonal) and form all 
two-electron states from these and diagonalize the Hamiltonian within this 
set, then, apart from the ground-state, the low-lying multiplet is a very 
poor approximation to the exact result, with level separations which can be 
many orders of magnitude too large. This also gives rise to spurious 
qualitative errors, such as the lifting the degeneracy of the two triplets 
at $B=0$. The reasons for this are as follows. The first two Hartree-Fock 
states are localized on opposite corners of a diagonal and are similar to 
the Hartree states. However, the next two Hartree-Fock states, which are localized 
near the other two diagonal corners of the square, are less localized due 
to the increased kinetic energy needed to ensure orthogonality. This 
increased width of the excited states gives rise to enhanced tunneling 
between the low-energy two-electron states, leading to a large increase in 
level separation since the tunneling matrix elements are very sensitive to 
the width of the localized states. Furthermore, as the symmetry of the square
geometry is lost degeneracies arising from this symmetry are lifted. 
Conversely, the lowest one-electron Hartree states, located near each corner 
of the square, are sufficiently localized to account for most of the Coulomb
and kinetic energy, whilst still maintaining this symmetry.\\

\begin{figure}
\centerline{\epsfxsize=100mm \epsfbox{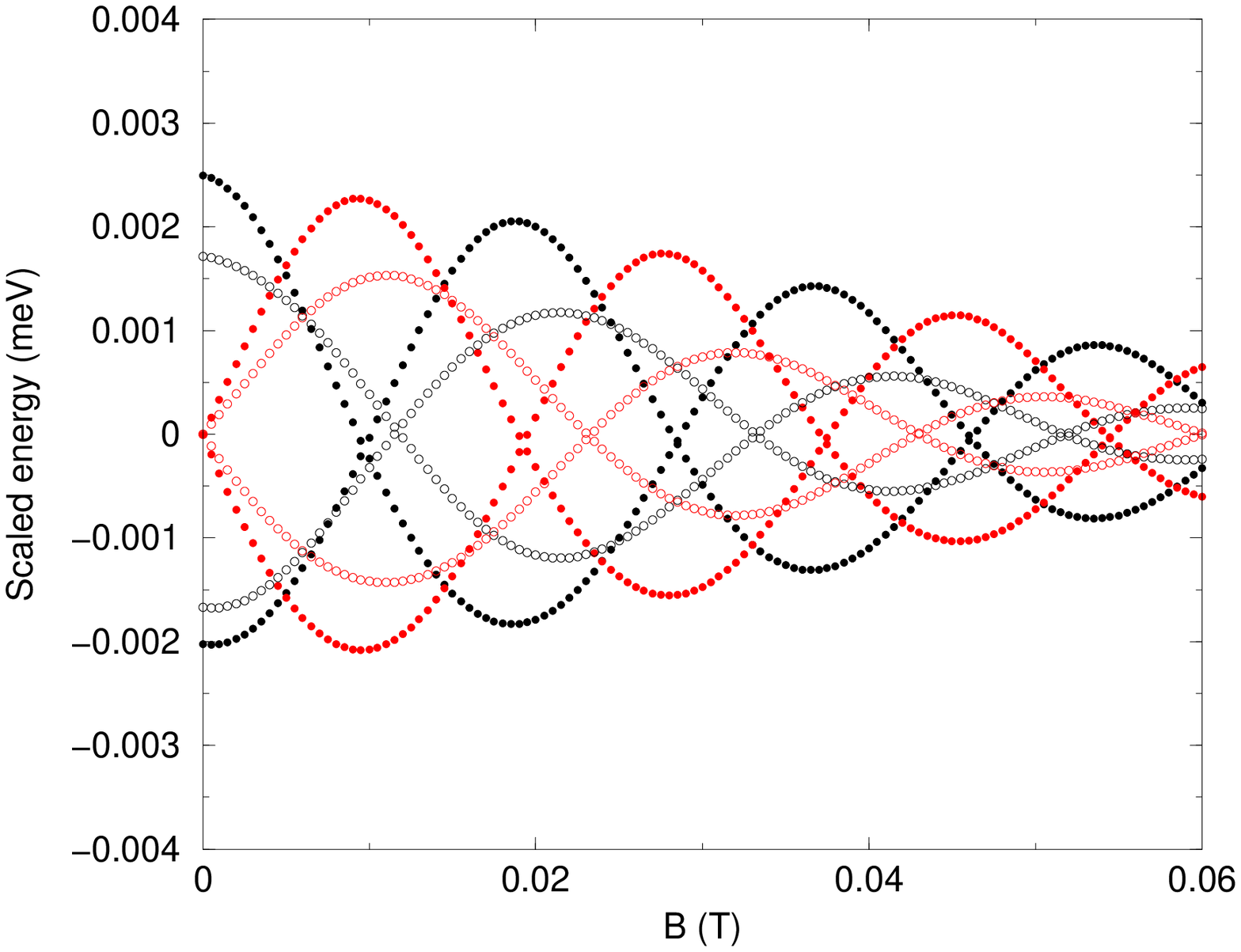}}
\centerline{\epsfxsize=100mm \epsfbox{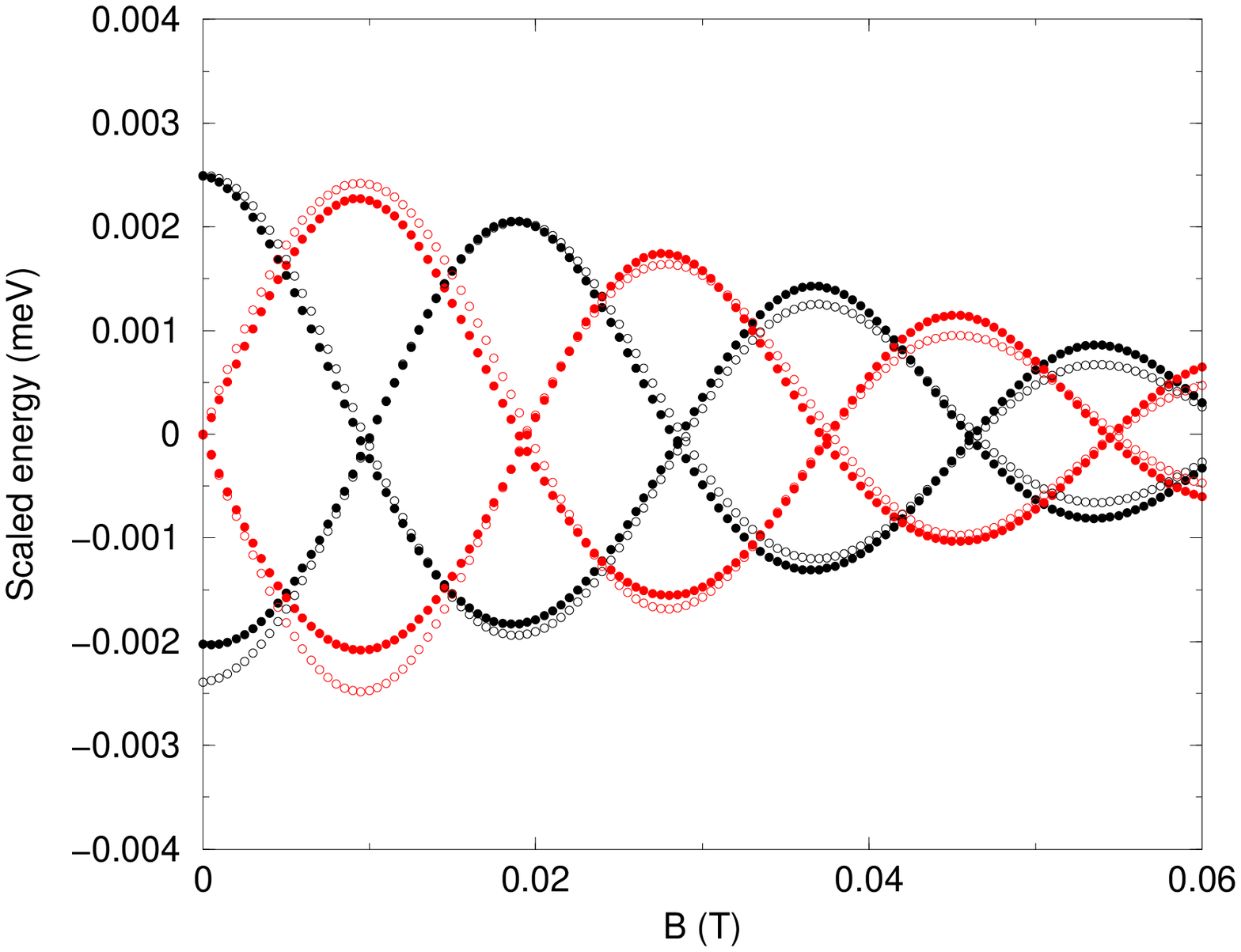}}
\caption{\label{tVapprox} Nearest neighbor $tV$-Hubbard model (open circles)
with kinetic hopping only (top) and with Coulomb-induced effective hopping
included (bottom). Exact results (filled circles) are shown for comparison.}
\end{figure}

We now consider in more detail the matrix elements of the Hamiltonian within
the basis set of the lowest orthogonalized Hartree states, and determine
which are essential to reproduce the exact results to good accuracy. If we
retain only the largest (diagonal) Coulomb matrix elements and the
one-electron matrix elements of the kinetic energy and confining potential,
then the second quantized Hamiltonian has precisely the form of 
Eq. (\ref{tv}) provided we preclude double occupation of a 
localized orbital, which is equivalent
to setting the intra-`site' Coulomb matrix element ($U$) to infinity. The
low-energy manifold within this approximation 
is plotted in Fig.[\ref{tVapprox}(a)], together with the exact solutions. 
We see that the main source of error is in the phase of
the oscillations, which have a significantly reduced frequency. Since we
have only neglected off-diagonal Coulomb matrix elements in this
approximation, the source of the error must arise from Coulomb-induced
effective hopping. There are many such terms, most of them small. One class
of terms which is not small involves three sites without spin-flips. The sum
of all such terms gives the following contribution to the effective
Hamiltonian: 
\begin{equation}
H_{{\rm 3site}}={\cal P}\sum\limits_{%
{\scriptstyle ijk\sigma \sigma ^{\prime }\hfill  \atop \scriptstyle i\neq j\neq k\hfill }%
}{\langle ik|g|jk\rangle c_{i\sigma }^{\dagger }}c_{k\sigma ^{\prime
}}^{\dagger }c_{k\sigma ^{\prime }}^{{}}c_{j\sigma }^{{}}{\cal P}={\cal P}%
\sum\limits_{%
{\scriptstyle ijk\sigma \hfill  \atop \scriptstyle i\neq j\neq k\hfill }%
}{\langle ik|g|jk\rangle c_{i\sigma }^{\dagger }}c_{j\sigma }^{{}}n_{k}{\cal %
P}={\cal P}\sum\limits_{\langle ij\rangle \sigma }{\left[ {%
t_{ij}^{C}c_{i\sigma }^{\dagger }c_{j\sigma }^{{}}+{\rm h}{\rm .c}{\rm .}}%
\right] }{\cal P} 
\label{threesite}
\end{equation}
where the restriction that all three sites be different follows from the
constraint of no double occupation, enforced by the projection operator, 
${\cal P}$; $t_{ij}^{C}=\langle ik|g|jk\rangle$, independent of $k$, and we
have used $P\sum_{k}n_{k}P=1$. Thus the effect of these Coulomb terms is to
simply renormalize the one-electron hopping $t_{ij}$. Plots of the
low-energy manifold are again shown in Fig.[\ref{tVapprox}(b)] and we see
that the correct Aharonov-Bohm oscillations are reproduced to good accuracy.
Thus we see that the simple nearest-neighbor $tV$-Hubbard model with 
$U=\infty$ describes accurately the low-energy physics provided that
the Coulomb contribution to the hopping is included.\\

\begin{figure}
\centerline{\epsfxsize=100 mm \epsfbox{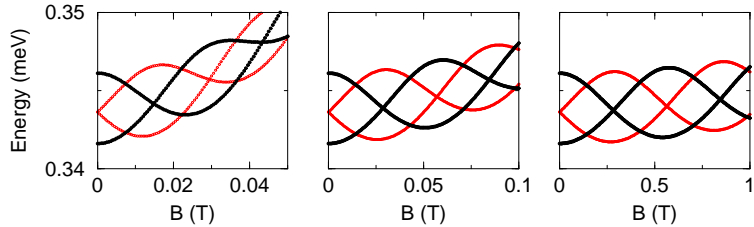}}
\caption{\label{fluxtube}
Energy levels of a two electron square-well quantum dot pierced by
a flux-tube. Flux-tube radii are: a) $L/2$; b) $L/4$; c) $L/16$.} 
\end{figure}

Within the $tV$-model we see that $B$-dependent terms arise in two different
ways. The Peierls factor, $e^{i\phi }$, comes solely from the magnetic flux
enclosed by the lattice, whereas the self-energy and the renormalization of
the $tV$ parameters originate from the physical interaction of the electrons
with the magnetic field. As we reduce the size of the flux-tube and approach
the limit of an Aharonov-Bohm flux-line, we should therefore expect the
latter effects to vanish, but for the Peierls factors to remain. To check
this supposition we display in Fig.[\ref{fluxtube}] the behavior of the
energy levels for flux-tubes of radius $L/2$, $L/4$ and $L/16$. It can be
clearly seen that as the radius of the flux-tube is reduced the overall
increase of all the energies with increasing field is reduced, corresponding
to the reduction in magnitude of the self-energy. The energy level
oscillations remain, as expected, and their amplitude decays less rapidly.
In the limit of an Aharonov-Bohm flux-line it is clear that the amplitude of
the oscillations will remain constant, and that no overall increase of 
energies will occur, meaning that the system can indeed be modeled by 
making a pure Peierls substitution in the $tV$ Hamiltonian.\\

The parallels between these results, and those obtained recently by Kotlyar 
{\it et al} [\cite{stafford}] for an array of coupled quantum dots holding a
small number of electrons, are striking. In the quantum dot arrays, electron
localization is achieved by confining electrons to individual quantum dots,
which are connected to their neighbors by leads. Exactly the same processes
of localization and hopping are present in the single quantum dot system we
consider, but localization comes from the electrostatic repulsion between
the electrons, which for sufficiently large dots forces the electrons into a
highly localized Wigner molecule state, and hopping occurs via tunneling
processes between the various low-lying energy levels. 
We note, however, that the fabrication of a single large quantum dot is
considerably simpler than the process of linking individual dots with leads,
and the creation of a dot with the dimensions we discuss is well within
current experimental capabilities. A single dot may thus provide a more
convenient physical realization to study strongly correlated mesoscopic
systems than an array of dots. In addition, within a single nanostructure it
should be possible to maintain coherence of the electrons for longer
time-scales, which is of relevance to the potential of these devices as
elements of quantum computers.\\ 

In conclusion, we have studied the behavior of the low-lying energy levels
of a square-well quantum dot containing two electrons, subject to a
perpendicular magnetic field. It has been shown in an earlier work that an
effective lattice model (a $tV$ model) can be used successfully to treat the
case of zero magnetic field, and we find that by making a simple Peierls
substitution this model also predicts the main qualitative changes to the
energy spectrum. This substitution gives exact results in the limit of an
Aharonov-Bohm flux-line, but to obtain quantitatively accurate results for
other flux distributions the parameters of the model must be magnetic-field
dependent, resulting in an approximately quadratic increase in average
energy with $B$, and a decrease in the amplitude of Aharonov-Bohm
oscillations. We have justified this behavior by employing a
single-electron Hartree basis in which the electrons are located close to
their electrostatic minima, near diagonally opposite corners of the square.
Within this framework, the overall increase in energy with magnetic field is
mainly due to the increase in the one-electron Hartree energy whilst the
reduction in amplitude of the Aharonov-Bohm oscillations is due to the
increased sharpness of the localized states, which reduces the
single-electron hopping. The Aharonov-Bohm oscillations are themselves a
consequence of the phase change when moving from one localized state to
another around the square, resulting in Peierls phase factors in the hopping
matrix elements and persistent currents around the perimeter of the square.
The system is thus equivalent to a tight-binding ring with four sites, the
quasi one-dimensionality being a consequence of Coulomb repulsion. Coulomb
repulsion also has the effect of forcing the electrons to be diagonally
opposite each other, causing them to rotate as a rigid pair, with an
Aharonov-Bohm oscillation of twice the frequency of non-interacting
electrons on a ring. The mapping to an effective lattice model is thus a
valuable way of interpreting the phenomena revealed in the spectrum of the
dot as the magnetic field is applied, and provides, for example, an
appealing interpretation of the origin of the parity oscillations of the
ground-state. This investigation has concentrated in the regime of weak to
medium strength magnetic fields, in which the magnetic length scale is
comparable with the Coulomb interaction length. 
Accordingly the spectrum shows a rich structure arising from the interplay
between magnetic effects and correlations produced by the Coulomb
interaction. As the magnetic field is increased further magnetic effects
will become dominant and the eigenstates of the system will evolve toward
Landau level states. Extending the range of magnetic field to study this 
transition, and its relation to the quantum Hall effect, is an exciting 
prospect for future investigation.\\

The authors would like to thank Wolfgang H\"ausler and Colin Lambert for
stimulating discussions. CEC acknowledges support from the Leverhulme
Foundation. Support from the UK Ministry of Defence and the EU TMR programme
is also acknowledged.

\bigskip

\end{document}